\documentclass[final,5p,times,twocolumn]{elsarticle}

\usepackage{amssymb}
\usepackage{amsmath}

\usepackage{graphicx}
\usepackage{color}
\usepackage{hyperref}
\usepackage{cuted}

\journal{Physica C}

\begin{document}

\begin{frontmatter}



\title{The Instability of a Quantum Superposition of Time Dilations}


\author[unige,lorentz]{Louk Rademaker} 
\author[leiden]{Tom van der Reep}
\author[leiden]{Nick Van den Broeck}
\author[leiden]{Bob van Waarde}
\author[leiden]{Marc de Voogd}
\author[leiden]{Tjerk Oosterkamp}
\affiliation[unige]{organization={Department of Quantum Matter Physics},
            addressline={24 Quai Ernest-Ansermet}, 
            city={Geneva},
            postcode={1211}, 
            country={Switzerland}}
\affiliation[lorentz]{organization={Institute-Lorentz for Theoretical Physics, Leiden University},
            addressline={PO Box 9506}, 
            city={Leiden},
            postcode={NL-2300}, 
            country={The Netherlands}}        
\affiliation[leiden]{organization={Leiden Institute of Physics},
            addressline={Niels Bohrweg 2}, 
            city={Leiden},
            postcode={2333 CA}, 
            country={The Netherlands}}

\begin{abstract}
Using the relativistic concept of time dilation, we show that a superposition of gravitational potentials can lead to nonunitary time evolution. For sufficiently weak gravitational potentials one can still define, for all intents and purposes, a global coordinate system. A probe particle in a superposition of weak gravitational fields will, however, experience dephasing due to the different time dilations. The corresponding instability timescale is accessible to experiments, and can be used as a degree of macroscopicity. We estimate this timescale for an experiment with smoothly tunable amplification in a microwave interferometer, that allows a quantitative study of the quantum to classical boundary.
\end{abstract}

\end{frontmatter}


\section{Introduction}

A core concept of general relativity is that inside a gravitational potential time runs slower. The relevance of this time dilation effect is by no means limited to astronomically large systems. A famous example is the need for gravitational time dilation corrections used in GPS positioning. With the relativistic influence penetrating at such small scales, one might wonder whether time dilation might influence quantum mechanical experiments.

Indeed, many proposals \cite{Lammerzahl,Dimopoulos,Zych,Pikovski2012,Pikovski2013} have been put forward that allow a quantum particle in spatial superposition to have different time evolutions depending on where it is located in the gravitational potential. Quantitatively, the time dilation arises from the $g_{00}$-component of the metric, so that in lowest order in the gravitational potential $\Phi(x)$ the local elapsed time $d\tau$ is
\begin{equation}
	d\tau = \left( 1 + \frac{\Phi(x)}{c^2} + \ldots \right) dt.
\end{equation}
This is less than the time interval $dt$ of the infinitely far away observer.\cite{WaldGR}

Here we are interested in the opposite question: \emph{what happens if we have a superposition of different gravitational potentials?}

Such a gravitational superposition naturally arises if one puts a massive object in a superposition. Formally, it is impossible to describe such a system since we can not make a unique mapping between two different space-times coordinates. Indeed, it has been suggested by Di\'{o}si\cite{Diosi} that the uncertainty associated with different space-time geometries will lead to a breakdown of the quantum mechanical superposition principle. This uncertainty is then characterized by a fluctuation noise term, a theme which has been studied substantially.\cite{Penrose1996,Adler,VanWezel2007,VanWezel2010,VanWezel2011,OZpaper} Several recent reviews give an excellent overview of the current status of gravitational effects on quantum mechanics.\cite{Bassi2013,Arndt2014,Bassi.2017,Anastopoulos.2022,Bose.2023}

Contrary to the Di\'{o}si assumption, in this publication we show that in the presence of weak gravitational superpositions a global coordinate system for space-time is well-defined for all intents and purposes. To study the coherence of the space-time coordinates we use a variation of Einstein's notion of a light clock in section \ref{SecLightClock}. Even though the space-time coordinates remain well-defined in this weak limit, the effects on quantum time evolution of a gravitational superposition will lead to nonunitary behavior. In section \ref{SecProbe} we derive an instability time for a probe particle in a gravitational superposition. To our surprise, the associated timescale
\begin{equation}
	\tau_p = \frac{\pi \hbar}{m_p |\Phi_1 (x_p) - \Phi_2(x_p)|}
\end{equation}
is within experimental reach. The instability times, corresponding to a number of existing experiments, are computed in section \ref{SecProbe}. In section \ref{SecExp} we discuss an experiment that allows for a dynamic detection of the instability of a gravitational superposition, using tunable amplification within a microwave interferometer \cite{Reep2021}. Above a critical gain, the interference pattern could disappear due to the aforementioned instability. Finally, we conclude this paper with a discussion of possible caveats and subtleties connected to our present theoretical and experimental proposals.

\section{Light clocks}
\label{SecLightClock}

As argued in the introduction, the very fabric of space-time becomes ill-defined once one allows superpositions of different space-times. This is true in the formal sense: there is no unique mapping between points in different space-times. However, imagine a ball with mass $M$ in superposition of having a small radius $a$ and a big radius $b$. Far outside the ball we can make a clear identification of the two space-times since at such distance the gravitational potential is independent of the ball's radius. Using a variation on Einsteins light-clock argument, we will show that close to the ball the time-coordinate is still well-defined for all practical purposes. 

\begin{figure}
	\includegraphics[width=\columnwidth]{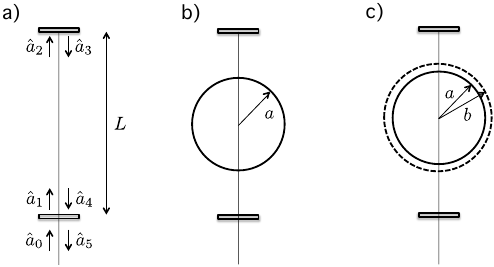}
	\caption{Three versions of a light clock, measuring the Shapiro delay due to a gravitational potential. \textbf{a}. In the absence of a gravitational potential, the light bounces off at the top mirror. \textbf{b}. A massive ball with radius $a$ and mass $M$ is placed in between the two mirrors, causing a Shapiro delay. \textbf{c.} If the massive ball is in a superposition of two different radii $a$ and $b$, the light travelling between the mirrors picks up different delays, causing the light clock to lose coherence.}
	\label{FigLightClock}
\end{figure}

Recall that a light-clock consists of a light pulse sent back and forth between two mirrors at a distance $L$ from each other. Without time dilation, the light pulse comes back every $2 \Delta t = \frac{2L}{c}$ to its original position, see Fig. \ref{FigLightClock}a.

Imagine there is, as in Fig. \ref{FigLightClock}b, a massive transparent ball with mass $M$ and radius $a$ in between the mirrors causing time dilation of the light pulse. The time it takes light to traverse the cavity once is 
\begin{equation}
	\Delta t = \frac{1}{c} \int_0^L dx \left( 1 - \frac{\Phi(x)}{c^2} \right)
\end{equation}
where we the integral over the gravitational potential of a solid sphere yields 
\begin{equation}
	\int_0^L dx \Phi(x)
		= 2 G M \left( \log \frac{2a}{L} - \frac{4}{3} \right)
\end{equation}
so that
\begin{equation}
	\Delta t = \frac{L}{c} + \frac{2 G M}{c^3} \left( \log \frac{L}{2a} + \frac{4}{3} \right).
\end{equation}

What happens now if we have the aforementioned superposition of different time dilations effects? Consider that the massive ball is put in a superposition of having radii $a$ and $b$, which is described as
\begin{equation}
	| \psi_{\mathrm{ball}} \rangle
		= \frac{1}{\sqrt{2}} \left( | a \rangle 
		+ | b \rangle \right),
\end{equation}
see Fig. \ref{FigLightClock}c. Notice that the gravitational potential outside the outer radius is the same for both states. A pulse sent in at a given time comes back in a superposition of three possible arrival times: back and forth with a delay caused by state $a$ or $b$, and back with a delay caused by $a$ and forth with a delay caused by state $b$. The maximal time difference between superposed arrival times is given by
\begin{equation}
	2\delta t = \frac{4GM}{c^3} \log \frac{b}{a}.
\end{equation}
This is an exceedingly small number: take for example a ball of mass $M = 10^{-12}$ g with a radius difference of $\frac{a}{b} = 0.95$. Then the time difference accumulated after one full reflection is $\delta t \sim 10^{-49}$ s, indeed seemingly irrelevant.

We can get a closer understanding of the functioning of the light clock by studying the quantum optics of the clock. Start with a light pulse with wavepacket amplitude $\xi_{\mathrm{in}}(t)$ at $\hat{a}_0$. In the absence of time dilation superpositions, Fig. \ref{FigLightClock}a, the light picks up a phase factor
\begin{equation}
	\hat{a}_2 = e^{i \omega \, \Delta t} \hat{a}_1
	\label{FieldTraverse}
\end{equation} 
when traversing the cavity once. The light is then reflected at the upper mirror, where we neglect the currently irrelevant reflection phase factor, hence $\hat{a}_3 = \hat{a}_2$. A relation similar to Eqn. (\ref{FieldTraverse}) holds between the fields $\hat{a}_4$ and $\hat{a}_3$. The half-silvered mirror at the bottom partially reflects the incident light pulse, described by $\hat{a}_1 = \mathcal{T} \hat{a}_0 + \mathcal{R} \hat{a}_4$. The transmitted part $\hat{a}_5$ is obtained from combining these relations, yielding
\begin{equation}
	\hat{a}_5 = \frac{ \mathcal{T} e^{i \omega \, 2 \Delta t}}
	{1 - \mathcal{R} e^{i \omega \, 2  \Delta t}} \hat{a}_0.
\end{equation}
Since $|\mathcal{T}| \ll 1$ and thus $|\mathcal{R}| \approx 1$, we find that the outgoing wavepacket amplitude is given by
\begin{equation}
	\xi_{\mathrm{out}}(t) \sim \sum_n 
	\xi_{\mathrm{in}} \left( t - n 2 \Delta t\right),
\end{equation}
that is, pulses with a delay of $2 \Delta t$ come out of the light clock. This is indeed what we expect from a light clock in a flat space-time.

In the case of a superposition of time dilations, as shown in Fig. \ref{FigLightClock}c, the field relation Eqn. (\ref{FieldTraverse}) should incorporate this superposition. Here we arrive at uncharted territory. However, we feel a logical choice to replace Eqn. (\ref{FieldTraverse}) would be to insist that the outgoing field displays a superposition of the two time dilations $\Delta t(a)$ and $\Delta t(b)$,
\begin{equation}
	\hat{a}_2 = \left( \frac{1}{2} e^{i \omega \, \Delta t(a)}+
	\frac{1}{2} e^{i \omega \, \Delta t(b)}
	\right) \hat{a}_1.
	\label{PhotonDephasing}
\end{equation} 
Note that the factors $\frac{1}{2}$ are required because if we would artificially make a superposition of two times the same state, it should reduce to Eqn. (\ref{FieldTraverse}). Another way to understand the factors $\frac{1}{2}$ is to make the light clock infinitesimally small, that is $\Delta t\rightarrow 0$, in which case we should have $\hat{a}_2=\hat{a}_1$. Following this reasoning, a superposition $\alpha | a\rangle + \beta | b \rangle$ would thus give rise to a phase factor $\left( |\alpha|^2 e^{i \omega \, dt(a)}+ |\beta|^2 e^{i \omega \, dt(b) }\right) $, and in general we would have
\begin{equation}
	\hat{a}_2 = \left( \sum_n |c_n|^2 e^{i \omega dt(n)} \right) \hat{a}_1
	\label{Dephasing3}
\end{equation}
if the massive object is in superposition $| \psi_{\mathrm{ball}} \rangle=\sum_n c_n |\psi_n\rangle$, where each $|\psi_n\rangle$ corresponds to a well-defined classical space-time and $\sum_n |c_n|^2=1$. The time dilation difference now causes destructive interference between the differently delayed light pulses. The magnitude of the destructive interference can be inferred from the outgoing wavepacket amplitude, which is approximately 
\begin{eqnarray*}
	\xi_{\mathrm{out}}(\omega) &\approx&
		\sum_n 
		 \frac{\xi_{\mathrm{in}}(\omega)}{4^n} e^{- i \omega 2n \overline{\Delta t}}
			\left( e^{i \omega \frac{\delta t}{2}}+
			e^{-i \omega \frac{\delta t}{2}} \right)^{2n} 
		 \\
	&=& \sum_{n} \sum_{k=0}^{2n} 
			\frac{\xi_{\mathrm{in}}(\omega)}{4^n}
			\binom{2n}{k} e^{- i \omega ( 2n \overline{\Delta t} + (n-k) \delta t)}
	 \\ 
	& \approx & \sum_n \xi_{\mathrm{in}}(\omega)
	\int \frac{dk \, e^{-\frac{(k-n)^2}{n}}}{\sqrt{\pi n}} 
	e^{- i \omega ( 2n \overline{\Delta t} + (n-k) \delta t)},	
\end{eqnarray*}
where we defined $\overline{\Delta t} = \frac{1}{2} \left( \Delta t(a) + \Delta t(b) \right)$ and $\delta t = \frac{1}{2} \left( \Delta t(a) - \Delta t(b) \right)$, and in the last line we have used the continuum expression for the binomial coefficient for large $n$. From the Fourier transform we extract the height of the arriving pulses, given that the initial pulse had the shape $\xi_{\mathrm{in}}(t) = e^{-\Delta \omega^2 t^2}$ where $\Delta \omega$ is the bandwidth of the pulse. We find, for large times $t = 2n\overline{\Delta t}$,
\begin{equation}
	\xi_{\mathrm{out}}(t )
	\sim \frac{1}{\Delta \omega \, \delta t \, \sqrt{ t / 2 \overline{\Delta t}}}.
\end{equation}
This implies that for our earlier result of $\delta t \sim 10^{-49}$ s, a typical bandwidth of an optical laser of $\Delta \omega \sim 10^{12}$ Hz (i.e. subpicosecond pulses), and a typical clock size such that $\overline{\Delta t} = 10^{-13}$ s, we need to wait $t = 10^{60}$ s before this clock will display appreciable deviations from running perfectly.

We have thus found that the time coordinate as measured by our light clock will behave properly, even in the presence of the weak superposition presented here. This suggests that we can still use the language of quantum mechanics, which requires a globally well-defined coordinate system, to describe superpositions of this kind. However, we must be looking for something heavier than a light-clock to see effects of the superposition of time dilations. 

\section{A massive probe particle}
\label{SecProbe}

Using the light clock, we obtained an idea how the existence of the gravitational superposition might influence the evolution of an external system. Eqn. (\ref{Dephasing3}) shows how we can write a superposition of time evolutions for a quantum wavepacket. In this section, we take this idea one step further and consider the quantum time evolution of a massive probe particle in such a gravitational potential.

In the absence of any potential, the time evolution of the state of a classical probe particle would be\footnote{Here we use the notion that a classical particle occupies a minimal uncertainty state, and therefore can be regarded as having a well-defined position $x_p$ and a well-defined energy $E$.}
\begin{equation}
	|\psi_p (t) \rangle = e^{-i E t/\hbar}	|\psi_{p0} \rangle.
\end{equation}
For any non-relativistic particle, the most of its energy will be contained in its rest-mass, hence $E=m_pc^2$ where $m_p$ is the mass of the particle. Note that this is again a concept from relativity, in quantum mechanics there is no absolute value of energy. We consider it to be a reasonable assumption to take the relativistic rest-mass energy of a particle.

\begin{figure}
	\includegraphics[width=\columnwidth]{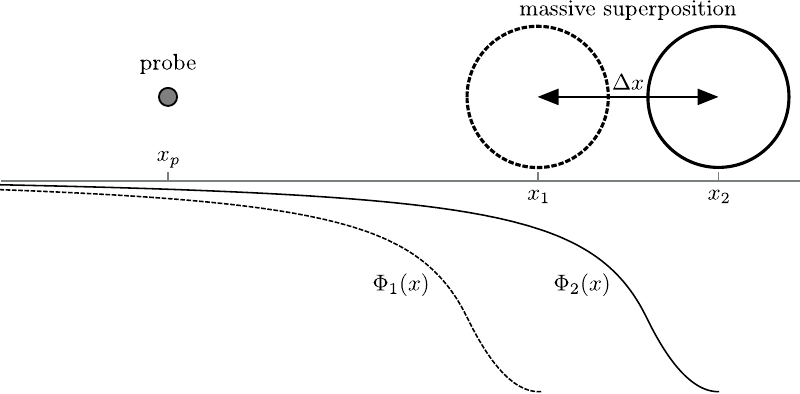}
	\caption{A heavy ball is brought into a superposition at positions $x_1$ and $x_2$. This leads to a superposition of the gravitational potentials $\Phi_1(x)$ and $\Phi(x)$. A nearby probe particle at position $x_p$ experiences a superposition of time evolutions.}
	\label{FigProbe}
\end{figure}

In the vicinity of a heavy particle the time dilation caused by its gravitational potential $\Phi(x)$ will slow down the phase of the probe particle. Assume the probe particle is located at position $x_p$, yielding a time dilation of
\begin{equation}
	|\psi_p (t) \rangle = e^{-i E t \left(
	1 + \frac{\Phi(x_p)}{c^2}
	\right) /\hbar}	|\psi_{p0} \rangle.
\end{equation}
The reader can anticipate the next step in our reasoning: what if the heavy object is in a superposition, as shown in Fig. \ref{FigProbe}?

Similar to the dephasing in our light clock, anticipated in Eqn. (\ref{PhotonDephasing}), we suggest a superposition of time evolutions
\begin{equation}
	\frac{1}{2} \left( e^{-i E t \left(
	1 + \frac{\Phi_1(x_p)}{c^2}
	\right) /\hbar} + e^{-i E t \left(
	1 + \frac{\Phi_2(x_p)}{c^2}
	\right) /\hbar} \right)	|\psi_{p0} \rangle
	\label{Superpos1}
\end{equation}
where $\Phi_1(x)$ and $\Phi_2(x)$ are the gravitational potentials of the ball located at $x_1$ and $x_2$, respectively. Similar to the light pulses discussed in the previous section, one notices a destructive interference given by this time evolution. In this specific case, we can argue that the superposed time evolution is unstable: if the accumulated phase from the dilation with $\Phi_1(x_p)$ differs from the phase due to $\Phi_2(x_p)$ by $\pi$, the state cannot be normalized. At this point we certainly know that standard quantum mechanics can no longer be applied because the state Eqn. (\ref{Superpos1}) becomes ill-defined! This occurs when
\begin{equation}
	\frac{Et}{\hbar c^2} \left( \Phi_1(x_p) - \Phi_2(x_p) \right) = \pm \pi.
\end{equation}
Using the earlier mentioned estimate for the probe particle energy $E=m_pc^2$, the time it takes for the state of the probe particle to become non-normalizable equals
\begin{equation}
	\tau_{p} = \frac{\pi \hbar}{m_p \left| \Phi_1(x_p) - \Phi_2(x_p) \right|}.
	\label{ProbeInstab1}
\end{equation}
We know that in the above presented case this time scale represents an assured breakdown of quantum mechanical evolution. However, in general it is rather an expression of the timescale over which nonunitary evolution becomes relevant, since it represents a timescale over which the norm of the probe particle state decreases.\footnote{In traditional quantum mechanics the norm of a state ket is irrelevant for the physics of a system. Our reasoning suggests that if one wants to reconcile quantum mechanics with relativity, the norm can become relevant.} With this interpretation, Eqn. (\ref{ProbeInstab1}) lies at the heart of our proposal. It allows for various generalizations, of which we wish to discuss two. First of all, instead of just one probe particle consider a collection of probe particles described by a mass density $\rho_p(x)$. 
The probe particle instability then occurs at the timescale
\begin{equation}
	\tau_{p} = \frac{\pi \hbar}{\int d^3x \rho_p(x) \left| \Phi_1(x) - \Phi_2(x) \right|}.
	\label{ProbeInstab}
\end{equation}
Note that this is not the only possible generalization, a point which will be discussed in the outlook.

The other generalization is towards more complicated superpositions, as mentioned before in Eqn. (\ref{Dephasing3}). Imagine the heavy ball is in a superposition
\begin{equation}
| \psi_{\mathrm{ball}} \rangle=\sum_n c_n |\psi_n\rangle
\end{equation}
where $|\psi_n\rangle$ all represent different gravitational configurations, each with a potential $\Phi_n(x)$. As a generalization of Eqn. (\ref{Superpos1}), we postulate that the time evolution of the probe particle is given by 
\begin{equation}
	\sum_n |c_n|^2 e^{-i m_p \Phi_n t / \hbar} | \psi_{p0} \rangle
\end{equation}
The norm of this state
\begin{equation}
	\sqrt{ \sum_{n,m} |c_n|^2 |c_m|^2 \cos \left( m_p (\Phi_n - \Phi_m) t /\hbar \right)}
\end{equation}
decreases over time, signalling possible nonunitary time evolution. If we expand this norm into second order in $t$, we find a typical timescale over which the nonunitary processes occur,
\begin{equation}
	\tau_p = \frac{\pi \hbar}{m_p \sqrt{2 \sum_{n,m} |c_n|^2 |c_m|^2 \left(\Phi_n - \Phi_m\right)^2 }}.
	\label{ProbeInstab3}
\end{equation}
Here we placed the factors $\pi$ and $2$ such that it reduces to Eqn. (\ref{ProbeInstab1}) in the case of a superposition of two states. To understand the use of this equation, imagine a heavy particle with mass $M$ delocalized in a cube of volume $L^3$. A probe particle with mass $m$ at a distance $a \gg L$ from the box will have an instability time proportional to $\tau_p~\sim~\frac{\hbar a^2}{GMmL}$. Indeed, our physical intuition that increasing delocalization is increasingly unstable is satisfied.

Note that we also might combine the two generalizations, to express the instability time for a distribution of probe particles due to a complicated superposition:
\begin{equation}
	\tau_p = \frac{\pi \hbar}{\int d^3x \rho_p(x) \sqrt{2 \sum_{n,m} |c_n|^2 |c_m|^2 \left(\Phi_n(x) - \Phi_m(x)\right)^2 }}
	\label{ProbeInstab4}
\end{equation}

The actual computation of an instability time is a relatively straightforward task. For example, consider a mass of $10^{-15}$~kg. Given that a ball of that weight has a typical size of $1$ $\mu$m, we will put it in a superposition of $\Delta x = 1 \mu$m. Imagine a probe particle one order lighter, so $m_p = 10^{-16}$~kg, and put it at $10$~$\mu$m distance. The corresponding timescale is $\tau_p \sim 10^3$~s, which is an accessible timescale.

The advantage of the various shapes of the instability time, Eqns. (\ref{ProbeInstab1})-(\ref{ProbeInstab}) and (\ref{ProbeInstab3})-(\ref{ProbeInstab4}), is that they can be applied to a number of existing and proposed experiments. Thus we are able to compare the 'macroscopicity'\cite{Nimmrichter2013} of each such experiment in terms of their instability time. The results are shown in Table \ref{TabSuperpos}.

\begin{table}[b]
\begin{tabular}{lr}
Experiment & Instability time $\tau_p$ \\
\hline
Buckyball interference (1999)\cite{Arndt1999} & $2 \times 10^8$ s \\
Sodium interference (1988)\cite{Keith1988} & $ 10^6$ s \\
PFNS8 interference (2011)\cite{Gerlich2011} & $3 \times 10^6$ s \\
Nanosphere of $10^7$ amu\cite{Romero-Isart2011} & $ 10^5$ s \\
Neutron interference (2002)\cite{Zawisky2002} & $2 \times 10^3$ s \\
OTIMA with $10^8$ amu object\cite{Nimmrichter2011} & $6 \times 10^1$ s \\
Membrane phonons (2011) \cite{Teufel2011} & $ 10^0$ s \\
Micromirror of $10^{16}$ amu \cite{BM2014} &  $10^{-1}$ s \\
\hline
Microwave MZ interference (Sec. \ref{SecExp}) & down to $10^{-6}$ s \\
\end{tabular}
\caption{\label{TabSuperpos}Computed instability times for a number of existing and proposed superposition experiments. This list is inspired by the macroscopicity measure proposed by Ref. \cite{Nimmrichter2013}. We find that the traditional double-slit experiments have considerably longer instability timescales than the superpositions of membrane phonons or micromirrors. Our experimental proposal of a microwave Mach-Zehnder interferometer with tunable amplification in both arms allows us to tune the instability time $\tau_p$, see Sec. \ref{SecExp}. }
\end{table}

Consider for example the famous interference of C$_{60}$ buckyballs\cite{Arndt1999}. The buckyballs with mass $10^{-24}$~kg traversed a SiN$_x$ grating with a 100-nm period. If we treat our silicon-nitride grating as the probe particle, with mass density $3 \times 10^{3}$~kg$/$m$^3$ thickness $100$~nm and height and width approximately 1 mm, we find a timescale of $\tau_p \sim 2 \times 10^8$~s. Since the buckyball under consideration flew for $6$~ms before hitting the detection screen, we find that our predicted instability lies beyond the scope of the buckyball experiment. The other timescales $\tau_p$ in Table \ref{TabSuperpos} are computed in a similar fashion. Notice that in such computations both the mass of the experimental set-up, which we treat as the probe particle, as well as the mass of the superposed object matter. The shortest instability times are obtained for the proposed superposition of a micromirror of $10^{16}$ amu.\cite{BM2014,Marshall2003}  

\section{A tunable experiment}
\label{SecExp}

Using quite general arguments, we suggested that there must be an instability in a system probing a gravitational superposition. This nonunitary evolution might be the cause for the much-sought barrier between quantum mechanics and the classical realm. We are well aware that Eqn. (\ref{ProbeInstab}) is certainly not the only way to include time dilation superpositions in quantum mechanics. However, we postpone a discussion of alternatives to the outlook and assume Eqn. (\ref{ProbeInstab}) indeed describes the timescale over which gravitational superpositions become unstable.

How would one proceed to actually measure this timescale? Following the computations that lead to Table \ref{TabSuperpos} we anticipate that interference similar to the OTIMA experiment should become unstable within microseconds if the mass of the interfered object exceeds $10^{-12}$~kg, the weight of an average human cell. Imagine we would perform this experiment and find no interference pattern. Is this because of the instability of the gravitational superposition? Or is the absence of interference due to our bad experimental skills? Unfortunately, one cannot tell in general. However, if there exists a unitary to nonunitary transition, this can be seen in an experiment where the external parameters remain the same yet the size of a superposition is increased continuously. 

\begin{figure}
	\includegraphics[width=\columnwidth]{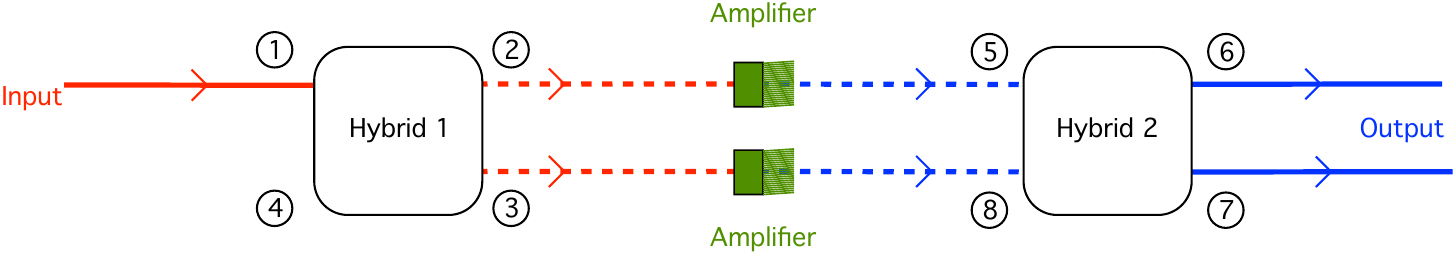}
	\caption{A microwave Mach-Zehnder interferometer with amplifiers in both arms.}
	\label{FigMZ2}
\end{figure}

Previously \cite{Reep2021}, we proposed an experiment entailing a Mach-Zehnder (MZ) interferometer with a single microwave photon source\cite{Bozyigit2011} in which we included a coherent amplifier to each of the interferometer arms, see fig. \ref{FigMZ2}. In such an experiment, a hybrid splits the single photons into an entangled superposition of being in the ``left'' and the ``right'' arm,
\begin{equation}\label{eqPsi0}
	| \psi \rangle = \frac{1}{\sqrt{2}} \left(
		|0 \rangle_L | 1 \rangle_R + |1 \rangle_L | 0 \rangle_R \right).
\end{equation}
In a normal MZ interferometer we bring the photon back to itself, and constructive or destructive interference arises depending on the path length difference of both arms.

The amplifier included in our proposal in both arms of the interferometer could be, for example, a degenerate traveling-wave parametric amplifier \cite{Vissers2016}. If the experiment is completely described by the laws of quantum mechanics, the amplification will lead to a state 
\begin{equation}\label{eqPsiAmp}
	|\psi\rangle = \frac{1}{\sqrt{2}} \hat{S}_L \hat{S}_R \left(
		\hat{a}^\dagger_L  + \hat{a}^\dagger_R  \right)
		|0\rangle,
\end{equation}
where $\hat{a}^\dagger_{L,R}$ is the creation operator for a photon in the left or right cable, respectively, and $\hat{S}$ is the squeezing operator that expresses the parametric amplification \cite{Loudon}. Interfering this state using the second hybrid, we found an amplifier-phase and -gain dependent visibility, maximizing to $\frac{1}{2}$ at high gain \cite{Reep2021}. 

The precise mechanism of state collapse was not under consideration in our proposal, although we did find deviations from the expected visibility in case state collapse sets in. In case the state collapses unto a number state, the interference visibility vanishes. On the other hand, if the state collapses unto a coherent state the visibility diminishes, but remains present \cite{Reep2021}. We expect that such a deviation might set in when we reach the timescale associated with the aforementioned instability. Let us now estimate the timescale over which this would happen.

To do so, we observe that microwave photons are carried by a electric potential wave $V(x,t)$ through a coaxial cable. Since both the electric potential and the gravitational potential are governed by Poisson's law, we relate the change in gravitational potential $\Phi_e(x)$ due to a microwave pulse to its electric potential $V(x)$,
\begin{equation}
	\Phi_e(x) = 4 \pi G \epsilon_0 \epsilon_r \frac{m_e}{|q_e|} V(x)
	\label{PhiV}
\end{equation}
where $\epsilon_r$ is the relative permittivity of the dielectric in the coaxial cable, $m_e$ is the mass of an electron and $q_e$ is the charge of an electron. We assume that the resistance and inductance of the outer conductor of the coaxial cable is so low that we can set the electric potential $V(x)$ zero outside the cable, and constant inside the cable.

 The classical states that correspond to a well-defined gravitational potential are thus states with well-defined voltage, which are the photon coherent states $|\alpha\rangle$. The voltage of a coherent state $| \alpha \rangle$ in the coaxial cable is given by
\begin{equation}
	V(x,t) = \sqrt{\frac{\hbar \omega_k}{2 \epsilon_0 \epsilon_r}}
		\; \alpha
		\; e^{-i \omega_k t - kx}.
	\label{Valpha}
\end{equation}
We should now express the quantum state after the degenerate parametric amplification, Eqn. (\ref{eqPsiAmp}), in the basis of coherent states. Using the completeness relation for coherent states we find the expansion
\begin{equation}
	|\psi \rangle
		= \int \frac{d^2\alpha_L d^2\alpha_R}{\pi^2}
		\langle 0 | \hat{D}^\dagger_L \hat{D}^\dagger_R | \psi \rangle
		\;
		| \alpha_L \otimes \alpha_R \rangle
	\label{Expansiona}
\end{equation}
in terms of the coherent states $\alpha_L$ and $\alpha_R$ in the left and right cable, respectively. Using this expansion, we use Eqn. (\ref{ProbeInstab4}) to compute the instability time as a function of the gain $\mathcal{G}$ of the amplifier in \ref{AppendixB}. We find that the instability time as a function of gain $\mathcal{G}$ equals
\begin{equation}
	\tau_p
	= \frac{1}{\sqrt{\mathcal{G}}}  \times 10^4 \, \mathrm{s}.
\end{equation}
Thus a single microwave photon without amplification should decay after roughly $10^4$ seconds. To arrive at microseconds, the time the pulse typically stays within the amplifier, we thus need a gain of about 200 dB. In order to obtain such large gains multiple amplifiers connected in series are necessary. However, our above analysis suggests that the visibility of the microwave interferometry should disappear at a critical gain of about 200 dB.

\section{Outlook}

In this paper, we have suggested that a superposition of different gravitational potentials can lead to an instability in the particles that experience this superposition. The key effect therein was time dilation. In this outlook, we will discuss the specific assumptions we made along the way. This should encourage future efforts in solving the problems of gravitational superpositions.

First of all, taking only time dilation into account seems a quite arbitrary choice. Even though the arguments from the light clock seem to support the idea that in the case of weak potentials we can still use one global flat coordinate system, an obvious rebuttal would be to question why we did not take into account space contraction. Space contraction, however, will not affect time evolution, and therefore we have neglected this effect in this paper.

Another major disputable assumption is the generalization from a single probe particle in Eqn. (\ref{ProbeInstab1}) towards a collection in Eqn. (\ref{ProbeInstab}). The converse is naturally correct, in that $\rho_p(x) = m_p \delta(x-x_p)$ equates the two timescales, but other generalizations might be possible as well. For example, a collection of point particles with mass $m_i$ at position $x_i$ can also have each their own instability time, so that the overall instability time is just the minimum of the collection,
\begin{equation}
	\tau_p = \min_i \left( \frac{\pi \hbar}{m_i \left| \Phi_1(x_i) - \Phi_2(x_i) \right|} \right).
\end{equation}
This form introduces problems of its own, such as how we should choose the size of the different point particles. We just introduce it here, to show that the equations used in this paper are by no means the only possible solution.

Furthermore, note that the concept of a probe instability is similar to standard decoherence theories, in that the coupling between the environment and the superposition leads to a collapse. However, as with decoherence theories, it is not trivial to decide which part of the environment one should use. Even if we take Eqn. (\ref{ProbeInstab}) for its face value, what is the correct choice for the probe mass density $\rho_p(x)$? Or what happens when the probe particle has a relative velocity with respect to the superposition? To the latter question, one might suggest a self-consistent equation of the form
\begin{equation}
	1 = \frac{\pi \hbar}{m_p \int^{\tau_p}_0 dt | \Phi_1 (x_p,t) - \Phi_2(x_p,t)|}
\end{equation}
which is Lorentz invariant: $\Phi' = \Phi / \gamma^2$, $t' = \gamma t$ and $m_p' = \gamma m_p$ under a Lorentz boost.\footnote{Imagine a particle in a gravitational potential $\Phi = gz$, such that the trajectory of this particle is $z(t) = z_0 - \frac{1}{2} gt^2$. A Lorentz boost in the $x$-direction yields $t' = \gamma t$ and thus $z'(t') = d - \frac{1}{2} g (t'/ \gamma)^2$. In the coordinate frame $(z',t')$ the particle thus experiences a gravitational potential $\Phi'=gz'/\gamma^2$.}

A subtle step we consciously avoided in this paper is the idea that the heavy particle in superposition might be its own probe, so that a completely isolated superposition will still have nonunitary evolution. Naively, we can suggest that the self-instability would occur at a timescale 
\begin{equation}
	\tau_p = \frac{\hbar \pi}{ M |\Phi_1(x_1) - \Phi_2(x_1)|} =  \frac{\hbar \pi}{ M |\Phi_1(x_2) - \Phi_2(x_2)|}
\end{equation}
where $M$ is the mass of the object in the state $|x_1 \rangle + |x_2 \rangle$. Whereas we certainly expect this to play a role in general, due to our uncertainties with regard to the specific expression for $\tau_p$ we think the simpler probe particle configuration needs to be understood in its entirety first. 

Let us also address the issue of entanglement. Consider a state where two heavy particles $a$ and $b$ are brought into an entangled superposition at the locations $a_1$ and $a_2$; and $b_1$ and $b_2$, respectively:
\begin{equation}
	|\psi_e\rangle = \frac{1}{\sqrt{2}} \left( |a_1 \rangle |b_1 \rangle
	+ |a_2 \rangle | b_2 \rangle \right)
\end{equation}
The associated timescale is now 
\begin{equation}
	\tau_{p,e} =\frac{\pi \hbar}{m_p \left| \Phi_{a1} - \Phi_{a2} + \Phi_{b1}  - \Phi_{b2} \right|}.
\end{equation}
Contrast this with the not-entangled superposition state
\begin{equation}
	|\psi_{ne}\rangle = 
	\frac{1}{2} \left( |a_1 \rangle +| a_2 \rangle \right)
	\left( | b_1 \rangle +| b_2 \rangle \right)
\end{equation}
which has a timescale
\begin{equation}
	\tau_{p,ne} = \frac{\pi \hbar}{m\sqrt{ (\Phi_{a1}-\Phi_{a2})^2 + (\Phi_{b1}-\Phi_{b2})^2 }}
\end{equation}
The two timescales are clearly different. When one assumes the probe particle is located at the origin and $a_1 < a_2$ and $b_1 < b_2$, the not-entangled state is more stable than the entangled state. However, when $a_1 < a_2$ and $b_2 < b_1$, the entangled state is more stable. It is interesting to investigate whether other theories of gravitational decoherence depend on whether a system is entangled or not.

Note that our analysis does not reproduce Born's rule nor the stochasticity associated with the measurement problem. Sometimes it is argued that Born's rule should be recovered by models that describe limitations on quantum superpositions, so that it can serve as an explanation of the measurement problem. The fact that our above model does not reproduce Born's rule, however, is not troublesome: the quantum coupling of the system to its macroscopic measurement apparatus that gives rise to Born's rule can be an independent effect, unrelated to a possible intrinsic instability of massive superpositions studied here.

Reading this outlook might have a discouraging effect on the reader since the number of open questions appears overwhelming. However, we feel that this paper overcomes two major obstacles we perceive earlier theories have: using time dilations we have shown that for weak potentials we can still use a global coordinate system; and secondly by isolating the effect of the gravitational superposition on a probe particle we find nonunitary evolution without noise or fluctuations. The instability we thus found can serve as a measure of macroscopicity that discerns the quantum and classical realm. At the same time, the experimental accessibility of our probe instability timescale by means of the proposed tuneable microwave interferometer suggests that the gravity-induced quantum to classical boundary is closer than was anticipated.

\section*{Acknowledgments}
The authors wish to thank Jasper van Wezel, Roger Penrose, Jan Zaanen, Martin van Exter and Gerard Nienhuis for fruitful discussions. This research is funded by the Netherlands Organisation for Scientific Research (NWO), through a Vici grant for TO and through a Rubicon grant for LR. LR is funded by the SNSF through Starting Grant TMSGI2\_211296. 

LR, TvdR and TO wrote the manuscript; LR, TO, MdV and BvW suggested the thought experiments; LR and TvdR analyzed the interferometer collapse timescale; and NVdB computed the values in Table \ref{TabSuperpos}.

\appendix

\section{Instability time in the interferometer}
\label{AppendixB}

We start with Eqn. (\ref{Expansiona}) where we express the state after parametric amplification in terms of coherent states. To obtain $\langle 0 | \hat{D}^\dagger_L \hat{D}^\dagger_R | \psi \rangle$, we first express the state $|\psi\rangle$ in terms of number states,
\begin{eqnarray}
	|\psi \rangle 
	&=& \frac{1}{\sqrt{2}}  \left(
		\hat{a}^\dagger_L \sqrt{\mathcal{G}}
		+ \hat{a}_L e^{-i \vartheta} \sqrt{1-\mathcal{G}}
		\nonumber \right. \\ && \left.
		+ \hat{a}^\dagger_R \sqrt{\mathcal{G}}
		+ \hat{a}_R e^{-i \vartheta} \sqrt{1-\mathcal{G}}
		 \right) \nonumber \\ &&
		\times \frac{1}{\sqrt{\mathcal{G}}}
		\sum_{m,n}
		\frac{\sqrt{(2n)!(2m)!}}{n!m!} 
		\left( -\frac{1}{2} e^{i \vartheta} \sqrt{\frac{\mathcal{G}-1}{\mathcal{G}}} \right)^{m+n}
		\nonumber \\ &&
		| 2m_L \otimes 2n_R \rangle.
\end{eqnarray}
Here $\mathcal{G}$ is the gain, $\vartheta$ is the phase of the parametric amplifier and $| m_L \otimes n_R\rangle$ is the state with $m$ photons in the left cable and $n$ photons in the right cable. Using the standard expansion of coherent states in terms of number states\cite{Loudon} we find the overlap $\langle 0 | \hat{D}^\dagger_L \hat{D}^\dagger_R | \psi \rangle$ equals
\begin{strip}
\begin{equation}
	\frac{1}{\mathcal{G}} \left( \alpha^*_L + \alpha^*_R \right)
		e^{-\frac{1}{2} (|\alpha_L|^2 + |\alpha_R|^2)
			- \frac{1}{2} ((\alpha^*_L)^2 + (\alpha_R^*)^2) e^{i\vartheta} \sqrt{\frac{\mathcal{G}-1}{\mathcal{G}}}  }.
\end{equation}
The term $\sum_{n,m} |c_n|^2 |c_m|^2 \left(\Phi_n(x) - \Phi_m(x)\right)^2$ inside the square root of the denominator of Eqn. (\ref{ProbeInstab4}) now becomes an integral over the coherent states of two different configurations $| \alpha_L \otimes \alpha_R \rangle$ and $|\alpha_L' \otimes \alpha_R' \rangle $,
\begin{eqnarray}
	\sum_{n,m} |c_n|^2 |c_m|^2 \left(\Phi_n(x) - \Phi_m(x)\right)^2&=&\int \frac{d^2 \alpha_L d^2 \alpha_R d^2 \alpha_L' d^2 \alpha_R'}{\pi^4}
	\frac{1}{\mathcal{G}^4} \nonumber \\ &&
	\left| \left( \alpha^*_L + \alpha^*_R \right)
		e^{-\frac{1}{2} (|\alpha_L|^2 + |\alpha_R|^2)
			- \frac{1}{2} ((\alpha^*_L)^2 + (\alpha_R^*)^2) e^{i\vartheta} \sqrt{\frac{\mathcal{G}-1}{\mathcal{G}}}  }
			\right|^2  \nonumber \\ &&
	\left| \left( (\alpha'_L)^* + (\alpha'_R)^* \right)
		e^{-\frac{1}{2} (|\alpha'_L|^2 + |\alpha'_R|^2)
			- \frac{1}{2} (((\alpha'_L)^*)^2 + ((\alpha'_R)^*)^2) e^{i\vartheta} \sqrt{\frac{\mathcal{G}-1}{\mathcal{G}}}  }
			\right|^2 \nonumber \\ &&
	\left( 4 \pi G \, \epsilon_0 \epsilon_r \, \frac{m_e}{|q_e|} \right)^2\; \frac{\hbar \omega_s}{2 \epsilon_0 \epsilon_r} 
	\left| \alpha_L - \alpha_L' \right|^2
\end{eqnarray}
where we have used Eqn. (\ref{PhiV}) to relate the gravitational potential to the electric potential and Eqn. (\ref{Valpha}) to relate the latter to the coherent state representation. Using
\begin{eqnarray}
	\int \frac{d^2 \alpha}{\pi}
	\;
	e^{-|\alpha|^2 - \frac{1}{2} ( (\alpha)^2 e^{-i\vartheta}+(\alpha^*)^2 e^{i \vartheta} \sqrt{\frac{\mathcal{G}-1}{\mathcal{G}}}}
	&=& \mathcal{G}^{1/2} \\
	\int \frac{d^2 \alpha}{\pi}
	\; |\alpha|^2 \;
	e^{-|\alpha|^2 - \frac{1}{2} ( (\alpha)^2 e^{-i\vartheta}+(\alpha^*)^2 e^{i \vartheta} \sqrt{\frac{\mathcal{G}-1}{\mathcal{G}}}}
	&=& \mathcal{G}^{3/2} \\
	\int \frac{d^2 \alpha}{\pi}
	\; |\alpha|^4 \;
	e^{-|\alpha|^2 - \frac{1}{2} ( (\alpha)^2 e^{-i\vartheta}+(\alpha^*)^2 e^{i \vartheta} \sqrt{\frac{\mathcal{G}-1}{\mathcal{G}}}}
	&=& 3 \mathcal{G}^{5/2},
\end{eqnarray}
and the fact that similar integrals with an odd power of $\alpha$ are zero, we find
\begin{equation}
	\sum_{n,m} |c_n|^2 |c_m|^2 \left(\Phi_n(x) - \Phi_m(x)\right)^2
	= 16 \mathcal{G}
	\left( 4 \pi G \, \epsilon_0 \epsilon_r \, \frac{m_e}{|q_e|} \right)^2\; \frac{\hbar \omega_s}{2 \epsilon_0 \epsilon_r} .
\end{equation}
\end{strip}
Finally, we need to insert this into Eqn. (\ref{ProbeInstab4}). We choose the  density $\rho_d = 10^4$ kg m$^{-3}$, $\epsilon_r = 20$ and frequency $f=10^9$ Hz $= 1 $ GHz. The coaxial cable is approximated as a cylinder of radius $r=1$ mm and length defined by the typical bandwidth of a microwave photon, thus $\ell=3$~m. Filling in the numbers yields an instability time of
\begin{equation}
	\tau_p
	= \mathcal{G}^{-1/2} 1.7 \times 10^4 \, \mathrm{s}.
\end{equation}

\end{document}